\documentclass[preprint,showpacs,preprintnumbers,amsmath,amssymb,groupaddress]{revtex4}
\usepackage{bbm}
\usepackage{amsfonts}

\usepackage{graphicx}
\usepackage{bm}

\begin{document}

\title{Application of the Hamiltonian formulation to nonlinear light-envelope propagations}
\author{Guo Liang }
\affiliation{Guangdong Provincial Key Laboratory of Nanophotonic Functional Materials and Devices, South
China Normal University, Guangzhou 510631, P. R. China}
\affiliation{School of Physics and Electrical Information, Shangqiu Normal University, Shangqiu 476000, P. R. China}
\author{Qi
Guo } \email{guoq@scnu.edu.cn} \affiliation{Guangdong Provincial Key Laboratory of Nanophotonic Functional Materials and Devices, South China Normal University, Guangzhou
510631, P. R. China}
\author{Yingbing Li }
\affiliation{Guangdong Provincial Key Laboratory of Nanophotonic Functional Materials and Devices, South
China Normal University, Guangzhou 510631, P. R. China}
\author{Zhanmei Ren}
\affiliation{Guangdong Provincial Key Laboratory of Nanophotonic Functional Materials and Devices, South
China Normal University, Guangzhou 510631, P. R. China}

\begin{abstract}
A new approach, which is based on the new canonical
equations of Hamilton found by us recently, is presented to analytically obtain the approximate solution of the nonlocal nonlinear Schr\"{o}dinger equation (NNLSE).
The approximate analytical soliton solution of the NNLSE can be obtained, and the stability of the soliton can be analytically analysed in the simple way as well, all of which are consistent with the results published earlier.  For the single light-envelope propagated in nonlocal nonlinear media modeled by the NNLSE, the Hamiltonian of the system
can be constructed, which is the sum of the generalized kinetic energy and the generalized potential. The extreme point of the
generalized potential corresponds to the soliton solution of the NNLSE. The soliton is stable when the generalized
potential has the minimum, and unstable otherwise. In addition, the rigorous proof of the equivalency between the NNLSE and the Euler-Lagrange equation is given on the premise of the response function with even symmetry.
\end{abstract}

.


\maketitle

\section{Introduction}
The propagations of the (1+D)-dimensional light-envelopes in nonlinear media
have been studied extensively for a few decades~\cite{Agrawal-book-01,Assanto-book-2012,Trillo-book-ss,Kivshar-book-os,Stegeman-science-1999,Chen-rpp-2012,Malomed-job-2005}, which are governed by the following dimensionless model~\cite{Guo-book-2015}:
\begin{equation}\label{NNLSE}
i\frac{\partial \varphi}{\partial
z}+\nabla_{\bot}^2\varphi+\Delta n\varphi=0,
\end{equation}
where $\varphi(\mathbf r,z)$ is the complex amplitude envelop, $\Delta n(\mathbf r,z)$ is the light-induced nonlinear refractive index, $z$ is the longitudinal coordinate, $\mathbf r$ is the
$D$-dimensional transverse coordinate vector with $D$ being the positive integer,
 and $\nabla_{\bot}$ is the $D$-dimensional differential operator vector of the transverse coordinates.
 Generally, $\Delta n(\mathbf r,z)$ can be phenomenologically expressed as the convolution between the response function $R(\mathbf r)$ of the media and the modulus square of the light-envelope $\varphi(\mathbf r,z)$ for the bulk media with the nonlocal nonlinearity~\cite{Guo-book-2015,Guo-book-2012,Snyder-science-97,Krolikowski-pre-01}
\begin{equation}\label{nonlocal-nonlinearity}
 \Delta n(\mathbf r,z)=\int_{-\infty}^{\infty} R(\mathbf r-\mathbf r^\prime)|\varphi(\mathbf r^\prime,z)|^2
\mathrm d^D\mathbf r^\prime.
\end{equation}
According to the relative scale of the
characteristic length of the response function {\it R} and the scale in the transverse dimension occupied by the light-envelope $\varphi$, the degree of nonlocality can be divided into four categories\cite{Guo-book-2015,Krolikowski-pre-01, Guo-book-2012}: local, weakly
nonlocal, generally nonlocal, and strongly nonlocal, and locality is the case when the response function {\it R} is the Dirac delta function.
In the local case, Eq.(\ref{NNLSE}) is reduced to
\begin{equation}\label{NLSE}
i\frac{\partial \varphi}{\partial
z}+\nabla_{\bot}^2\varphi+|\varphi|^2\varphi=0.
\end{equation}
Eq.~(\ref{NNLSE}) together with the nonlocal nonlinearity~(\ref{nonlocal-nonlinearity}) is called as the nonlocal nonlinear
Schr\"{o}dinger equation (NNLSE)~\cite{Guo-book-2015,Guo-book-2012,Krolikowski-pre-01}, while
 its special case, Eq.~(\ref{NLSE}), is the well-known nonlinear Schr\"{o}dinger equation (NLSE)~\cite{Kivshar-book-os,Agrawal-book-01,Trillo-book-ss}.

 The NNLSE (with its special case NLSE) can describe the nonlinear propagations of the optical beams~\cite{Assanto-book-2012,Trillo-book-ss,Stegeman-science-1999,Chen-rpp-2012,Kivshar-book-os}, the optical pulses~\cite{Agrawal-book-01,Kivshar-book-os} and the optical pulsed beams~\cite{Kivshar-book-os,Malomed-job-2005}. The second term of the NNLSE accounts for the diffraction  for the first case where $\mathbf r$ is the spatial transverse coordinate, the group velocity dispersion (GVD) for the second case where $\mathbf r$ is the time coordinate, and both the diffraction and the GVD  for the last case where $\mathbf r$ is both the spatial transverse coordinate and the time coordinate, while the third term (the nonlinear term) describes the compression of the light-envelopes for all cases. Specifically, when $D=1$, the NNLSE can model the propagation of the optical beam~\cite{Snyder-science-97,Krolikowski-pre-01} in the self-focusing nonlinear planar waveguide, and can also model the propagation of the optical pulse~\cite{Agrawal-book-01} in the self-focusing nonlinear waveguide if the carrier frequency is in the anomalous GVD regime or in the self-defocusing nonlinear waveguide when its carrier frequency is in the normal GVD regime. The (1+1)-dimensional NNLSE has the spatial (or temporal) bright optical soliton solution~\cite{Kivshar-book-os}. When $D=2$, the NNLSE can only describe the propagation of the optical beam in the nonlinear bulk media. The bright spatial optical soliton can exist stably for the nonlocal case~\cite{Guo-book-2012}, but for the local case the strong self-focusing of a two dimensional beam will lead to the catastrophic phenomenon~\cite{Kivshar-pr-2000}. When $D=3$, the NNLSE  can describe the propagation of the optical pulsed beams. Like the case of $D=2$, the self trapped optical pulsed beam propagating in the local nonlinear media will lead to the spatiotemporal collapse~\cite{Silberberg-ol-90}, which can be arrested by the nonlocal nonlinearity~\cite{Malomed-job-2005}. But when $D>3$, the NNLSE is just a phenomenological model, the counterpart of which can not be found in physics. It's important to note that\cite{Guo-book-2015} the response function $R$ is symmetric for the spatial nonlocality, but is asymmetric for the temporal nonlocality due to the causality~\cite{Hong-pra-2015}.

As the special case of the NNLSE, the NLSE (\ref{NLSE}) can be solved exactly using inverse-scattering technique~\cite{zakharov-zetf-1971,zakharov-zetf-1973} when $D=1$. But for the general case, a closed-form solution of NNLSE (\ref{NNLSE}) cannot been found except for the strongly nonlocal limit, where the NNLSE can be simplified to the (linear) Snyder-Mitchell model for the spatial nonlocality and an exact Gaussian-shaped stationary solution known as accessible soliton was found~\cite{Snyder-science-97}. Approximately analytical solutions can be obtained by various of perturbation methods, such as the perturbation approach based on the inverse scattering transform~\cite{Karpman-spj-1977}, the adiabatic perturbation approach~\cite{Kivshar-rmp-1989}, the method of moments~\cite{Maimistov-jetp-1993}, and the most widely used one is variational method~\cite{Anderson-pra-83,Malomed-progress in optics-2002,Guo-oc-06,chen-ol-2013}. It was claimed without proof that the variational method can only be applied in nonlocal cases where the response function is symmetric~\cite{Steffensen-josab-2012}. But the equivalency between the NNLSE (\ref{NNLSE}) and the Euler-Lagrange equation is not proved rigorously until the mathematical proof given in the paper on the premise of the response function with even symmetry. And for the case of the response function without even symmetry, the method of moments can work well. Another new approach is presented here, and
we apply the canonical equations of Hamilton to study the nonlinear light-envelope propagations. By taking this approach, the approximate analytical soliton solution of the NNLSE is obtained.
Furthermore, the stability of solutions can be analysed analytically in a simple way as well, but it can not be done by the variational approach.

The paper is organized as follows. We firstly give the rigorous proof of the equivalency between the NNLSE and the Euler-Lagrange equation in Sec. \ref{limition on response}, which is the basis of the variational approach applied in the NNLSE. The canonical equations of Hamilton (CEH) is a parallel method to the Euler-Lagrange equation in classical mechanics. But we find that the conventional CEH can not restate the NNLSE, and present a new CEH to restate the NNLSE, which is outlined in Sec.~\ref{ceh for 1st system}.
Based on the new CEH, we introduce a new approach in Sec.~\ref{application} to deal with the nonlinear light-envelope propagations. In Sec.~\ref{remarks} two remarks on the new approach are made. Firstly, we show that the conventional CEH will yield contradictory and inconsistent results. Secondly, we discuss the differences between our approach and the variational approach. Sec. \ref{conclusion} gives the summary.


\section{proof of the equivalency between the nonlocal nonlinear Schr\"{o}dinger equation and the Euler-Lagrange equation}\label{limition on response}
The variational approach is a widely used method to obtain the approximately analytical solution of the NLSE~\cite{Anderson-pra-83,Malomed-progress in optics-2002}.
The reason why the variational approach can be used is that the NLSE can be restated by the Euler-Lagrange equation, which reads (for the sake of simpleness, only
the case that D = 1 is taken consideration here)
\begin{equation}\label{euler-lagrange equation}
\frac{\partial}{\partial z}\frac{\partial l}{\partial\left(\frac{\partial\varphi^*}{\partial z}\right)}+\frac{\partial}{\partial x}\frac{\partial l}{\partial\left(\frac{\partial\varphi^*}{\partial x}\right)}-\frac{\partial l}{\partial\varphi^*}=0,
\end{equation}
if the
Lagrangian density $l$ is given by~\cite{Anderson-pra-83}
\begin{equation}\label{Lagrangian density for NLSE}
l=\frac{i}{2}\left(\varphi^*\frac{\partial\varphi}{\partial
z}-\varphi\frac{\partial \varphi^*}{\partial
z}\right)-\left|\frac{\partial\varphi}{\partial x}\right|^2+\frac{1}{2}\left|\varphi\right|^4.
\end{equation}
Replacing $\varphi^*$ with $\varphi$, the complex-conjugate equation of the NLSE can be obtained from the Euler-Lagrange equation (\ref{euler-lagrange equation}) consistently. Although the variational approach has been applied to the problems associated with the NNLSE (\ref{NNLSE}),
in which the Lagrangian density is expressed as~\cite{Guo-oc-06,chen-ol-2013}
 \begin{equation}\label{Lagrangian density for NNLSE}
l=\frac{i}{2}\left(\varphi^*\frac{\partial\varphi}{\partial
z}-\varphi\frac{\partial \varphi^*}{\partial
z}\right)-\left|\nabla_\bot\varphi\right|^2+l_{nl}
\end{equation}
with $l_{nl}\equiv\frac{1}{2}|\varphi(\mathbf r,z)|^2\Delta n(\mathbf r,z)$,
 the equivalency between the NNLSE (\ref{NNLSE}) and the Euler-Lagrange equation has not been proved rigorously. Without loss of generality, here we only give the proof of the equivalency in the case that $D=1$, and the extension to the general case of $D$ will be easy in a similar way.

Comparing the two expressions of the Lagrangian density for the NLSE and the NNLSE, i.e., Eqs.~(\ref{Lagrangian density for NLSE}) and (\ref{Lagrangian density for NNLSE}), we can observe that the Lagrangian density for the NNLSE contains a convolution between the response function $R$ and the modulus square of the light-envelope $\varphi$. Therefore, it has been somewhat confused how to calculate such terms as $\partial l_{nl}/\partial\varphi^*$ and $\partial l_{nl}/\partial\varphi$ for the NNLSE since $ l_{nl}$ is not the function of $\varphi$ and $\varphi^*$ but the functional of them. To this end, we first construct a functional as
\begin{eqnarray}
F(\varphi,\varphi^*)&\equiv&\int_{-\infty}^{\infty}\int_{-\infty}^{\infty}l_{nl}dxdz\nonumber\\
&=&\frac{1}{2}\int_{-\infty}^{\infty}\int_{-\infty}^{\infty}\int_{-\infty}^{\infty}R(x-x')|\varphi(x',z)|^2|\varphi(x,z)|^2dx'dxdz.
\end{eqnarray}
The variation of the functional $F(\varphi,\varphi^*)$ is defined as~\cite{Brizard-book-2008-1}
\begin{eqnarray}
\delta F(\varphi,\varphi^*)&=&\frac{\partial}{\partial \varepsilon}F(\varphi+\varepsilon\delta\varphi,\varphi^*+\varepsilon\delta\varphi^*)|_{\varepsilon\rightarrow0}\nonumber\\
&=&\frac{1}{2}\int_{-\infty}^{\infty}\int_{-\infty}^{\infty}\int_{-\infty}^{\infty}R(x-x')|\varphi(x,z)|^2\left[\varphi(x',z)\delta\varphi^*(x',z)+\varphi^*(x',z)\delta\varphi(x',z)\right]dx'dxdz\nonumber\\
&&+\frac{1}{2}\int_{-\infty}^{\infty}\int_{-\infty}^{\infty}\Delta n\left[\varphi(x,z)\delta\varphi^*(x,z)+\varphi^*(x,z)\delta\varphi(x,z)\right]dxdz.
\end{eqnarray}
If the response function is symmetric, i.e., $R(x)=R(-x)$, then we can obtain that
\begin{eqnarray}
&&\int_{-\infty}^{\infty}\int_{-\infty}^{\infty}R(x-x')|\varphi(x,z)|^2\left[\varphi(x',z)\delta\varphi^*(x',z)+\varphi^*(x',z)\delta\varphi(x',z)\right]dx'dx\nonumber\\
&&=\int_{-\infty}^{\infty}\Delta n\left[\varphi(x,z)\delta\varphi^*(x,z)+\varphi^*(x,z)\delta\varphi(x,z)\right]dx.
\end{eqnarray}
Then the variation of the functional $F(\varphi,\varphi^*)$ is simplified to
\begin{equation}\label{deal with convolution}
\delta F(\varphi,\varphi^*)=\int_{-\infty}^{\infty}\Delta n\varphi^*(x,z)\delta\varphi(x,z)dxdz+\int_{-\infty}^{\infty}\int_{-\infty}^{\infty}\Delta n\varphi(x,z)\delta\varphi^*(x,z)dxdz.
\end{equation}
On the other hand, the variation of the functional $F(\varphi,\varphi^*)$ can be also expressed as~\cite{Brizard-book-2008-2}
\begin{eqnarray}
\delta F(\varphi,\varphi^*)&=&\delta\int_{-\infty}^{\infty}\int_{-\infty}^{\infty}l_{nl}(\varphi,\varphi^*)dxdz\nonumber\\
&=&\int_{-\infty}^{\infty}\int_{-\infty}^{\infty}\left[\frac{\partial l_{nl}}{\partial\varphi}-\frac{\partial}{\partial z}\frac{\partial l_{nl}}{\partial\left(\frac{\partial\varphi}{\partial z}\right)}-\frac{\partial}{\partial x}\frac{\partial l_{nl}}{\partial\left(\frac{\partial\varphi}{\partial x}\right)}\right]\delta\varphi dxdz\nonumber\\
&&+\int_{-\infty}^{\infty}\int_{-\infty}^{\infty}\left[\frac{\partial l_{nl}}{\partial\varphi^*}-\frac{\partial}{\partial z}\frac{\partial l_{nl}}{\partial\left(\frac{\partial\varphi^*}{\partial z}\right)}-\frac{\partial}{\partial x}\frac{\partial l_{nl}}{\partial\left(\frac{\partial\varphi^*}{\partial x}\right)}\right]\delta\varphi^*dxdz\nonumber\\
&=&\int_{-\infty}^{\infty}\int_{-\infty}^{\infty}\frac{\partial l_{nl}}{\partial\varphi}\delta\varphi dxdz+\int_{-\infty}^{\infty}\int_{-\infty}^{\infty}\frac{\partial l_{nl}}{\partial\varphi^*}\delta\varphi^* dxdz.\label{variation}
\end{eqnarray}
Comparing Eqs.(\ref{deal with convolution}) and (\ref{variation}), we obtain
\begin{eqnarray}
\frac{\partial l_{nl}}{\partial \varphi^*}&=&\Delta n\varphi(x,z),\label{tonnlse}\\
\frac{\partial l_{nl}}{\partial \varphi}&=&\Delta n\varphi^*(x,z).\label{tonnlsecm}
\end{eqnarray}
 Inserting the Lagrangian density (\ref{Lagrangian density for NNLSE}) for the case of $D=1$ into the Euler-Lagrange equation (\ref{euler-lagrange equation}), the first two terms of (\ref{euler-lagrange equation}) can be easily obtained as
 \begin{equation}\label{first two terms}
 \frac{\partial}{\partial x}\frac{\partial l}{\partial\left(\frac{\partial\varphi^*}{\partial x}\right)}+\frac{\partial}{\partial z}\frac{\partial l}{\partial\left(\frac{\partial\varphi^*}{\partial z}\right)}=i\frac{\partial \varphi}{\partial
z}+\frac{\partial^2\varphi}{\partial x^2}.
 \end{equation}
Then the NNLSE (\ref{NNLSE}) can be obtained from the  Euler-Lagrange equation (\ref{euler-lagrange equation}) by using Eqs.~(\ref{tonnlse}) and (\ref{first two terms}) respectively, and its complex-conjugate equation can also be obtained consistently in a similar way.

Consequently, the NNLSE (\ref{NNLSE}) is equivalent to the Euler-Lagrange equation (\ref{euler-lagrange equation}) if the response function is symmetric. But for the asymmetric response function, for example, the response function
for the temporal nonlocality~\cite{Hong-pra-2015},
we can not show the equivalency between the NNLSE (\ref{NNLSE}) and the Euler-Lagrange equation (\ref{euler-lagrange equation}) anymore. In our points of view, the conclusion obtained here is equivalent to that given
in Ref.~\cite{Steffensen-josab-2012}, where the authors claimed that 
the equation (1) in Ref.~\cite{Steffensen-josab-2012} (similar to the NNLSE) does not have a Lagrangian when the temporally asymmetric nonlocal term is included and that ``Had the nonlocality been symmetric, then variational techniques could have been applied'', although no any proof was given in Ref.~\cite{Steffensen-josab-2012}. 

\section{Canonical equations of Hamilton for the NNLSE}\label{ceh for 1st system}
As discussed in the section above, the variational approach to find the approximately analytical solution of the NNLSE is based on the Euler-Lagrange equations. In the classical mechanics, however, there exist two theory frameworks: the Lagrangian formulation (the Euler-Lagrange equations) and the Hamiltonian formulation (canonical equations of Hamilton). The two methods are parallel, and no one is particularly superior to the another for the direct solution of mechanical problems~\cite{Goldstein-book-05}.
The new approach presented in this paper to analytically obtain the approximate  solution of the
NNLSE is based on the new
 canonical equations of Hamilton (CEH) found by us recently~\cite{liang-arxiv-2013}. For the sake of the systematicness
 and the readability of this paper, the key points about the new
CEH are outlined here in this section, although the detail can be found in Ref~\cite{liang-arxiv-2013}.

We firstly define two different systems of mathematical physics: the second-order differential system (SODS) and the first-order differential system (FODS).
The SODS is defined as the system described by the second-order partial differential equation about the evolution coordinate, while the FODS is defined as the system described by the first-order partial differential equation about the evolution coordinate. The Newton's second law of motion and the NNLSE are the exemplary SODS and FODS, where the evolution coordinates are the time coordinate $t$ and the propagation coordinate $z$, respectively.
The conventional CEH~\cite{Goldstein-book-05}
\begin{eqnarray}
\dot{q_i}&=&\frac{\partial H}{\partial p_i},\label{qi}\\
 -\dot{p_i}&=&\frac{\partial H}{\partial q_i},\label{pi}
\end{eqnarray}
is established on the basis of the Newton's second law of motion.
The dot above the variable in Eqs.~(\ref{qi}) and (\ref{pi}) ($\dot{q_i}$ and $\dot{p_i}$) indicates the derivative with respect to the evolution coordinate (here the evolution coordinate is the time $t$),
$q_i$ and $p_i$ are said to be the generalized coordinate and the generalized momentum, and $H$ is the Hamiltonian. The CEH (\ref{qi}) and (\ref{pi}) can be extended
to the continuous system as~\cite{Goldstein-book-05}
\begin{eqnarray}
\dot{q}_s&=&\frac{\delta h}{\delta p_s},\label{derivative of q for continuous system}\\
-\dot{p}_s&=&\frac{\delta h}{\delta q_s},\label{derivative of pi for continuous system}
\end{eqnarray}
with $s=1,\cdots,N$ representing the components of the quantity of the continuous system~\cite{Goldstein-book-05}, $\frac{\delta h}{\delta q_s}=\frac{\partial h}{\partial q_s}-\frac{\partial}{\partial x}\frac{\partial h}{\partial q_{s,x}}$ and $\frac{\delta h}{\delta p_s}=\frac{\partial h}{\partial p_s}-\frac{\partial}{\partial x}\frac{\partial h}{\partial p_{s,x}}$ denote the functional derivatives of $h$ with respect to $q_s$ and $p_s$ with $q_{s,x}=\frac{\partial q_s}{\partial x}$ and $p_{s,x}=\frac{\partial p_s}{\partial x}$, and $h$ is the Hamiltonian density of the continuous system.

We have shown that the FODS can not be expressed by the conventional CEH, and we have re-constructed  a set of new CEH through the following procedure.

For the first-order differential system of the continuous systems, the Lagrangian density must be the linear function of the generalized velocities, and expressed as
\begin{equation}\label{Lagrangian density for the fist-order system}
l=\sum_{s=1}^NR_s(q_s)\dot{q}_s+Q(q_s,q_{s,x}),
\end{equation}
where $R_s$ is not the function of a set of $q_{s,x}$ with $q_{s,x}=\partial q_s/\partial x$. Consequently, the generalized momentum $p_s$, which is obtained by the definition $p_s=\partial l/\partial\dot{q}_s$ as
\begin{equation}\label{equations of p q}
p_s=R_s(q_s), (s=1,\cdots,N),
\end{equation}
is only a function of $q_s$. There are $2N$ variables, $q_s$ and $p_s$, in Eqs. (\ref{equations of p q}). The number of Eqs. (\ref{equations of p q}) is $N$, which also means there exist $N$ constraints between $q_s$ and $p_s$. So the degree of freedom of the system given by Eqs. (\ref{equations of p q}) is $N$. Without loss of generality, we take $q_1,\cdots,q_\nu$ and $p_1,\cdots,p_\mu$ as the independent variables, where $\nu+\mu=N$. The remaining generalized coordinates and generalized momenta can be expressed with these independent variables as
$
q_\alpha=q_\alpha(q_1,\cdots,q_\nu,p_1,\cdots,p_\mu)(\alpha=\nu+1,\cdots,N),$ and $
p_\beta=p_\beta(q_1,\cdots,q_\nu,p_1,\cdots,p_\mu) (\beta=\mu+1,\cdots,N).
$
The Hamiltonian density $h$ for the continuous system is obtained by the Legendre transformation
as
$
 h=\sum_{s=1}^N\dot{q}_sp_s-l
$, where the Hamiltonian density $h$ is a function of $\nu$ generalized coordinates, $q_1,\cdots,q_\nu$, and $\mu$ generalized momenta, $p_1,\cdots,p_\mu$. We can obtain the new CEH consisting of N equations as
{\setlength\arraycolsep{0pt}
\begin{eqnarray}
\frac{\delta h}{\delta q_\lambda}&=&\sum_{s=1}^N\left(\dot{q}_s\frac{\partial p_s}{\partial q_\lambda}-\dot{p}_s\frac{\partial q_s}{\partial q_\lambda}\right)+\sum_{\alpha=\nu+1}^{N}\frac{\partial}{\partial x}\frac{\partial h}{\partial q_{\alpha,x}}\frac{\partial q_\alpha}{\partial q_\lambda},\label{canonical equations 1 for 1st order }\\
\frac{\delta h}{\delta p_\eta}&=&\sum_{s=1}^N\left(\dot{q}_s\frac{\partial p_s}{\partial p_\eta}-\dot{p}_s\frac{\partial q_s}{\partial p_\eta}\right)+\sum_{\alpha=\nu+1}^{N}\frac{\partial}{\partial x}\frac{\partial h}{\partial q_{\alpha,x}}\frac{\partial q_\alpha}{\partial p_\eta}\label{canonical equations 2 for 1st order },
\end{eqnarray}
}where $\lambda=1,\cdots,\nu$, $\eta=1,\cdots,\mu$, and $\nu+\mu=N$.
The CEH (\ref{canonical equations 1 for 1st order }) and (\ref{canonical equations 2 for 1st order }) can be easily extended to the discrete system, which can be expressed as
{\setlength\arraycolsep{0pt}
\begin{eqnarray}
\frac{\partial H}{\partial q_\lambda}&=&\sum_{s=1}^N\left(\dot{q}_s\frac{\partial p_s}{\partial q_\lambda}-\dot{p}_s\frac{\partial q_s}{\partial q_\lambda}\right),\label{canonical equations 1 for 1st order for discrete system }\\
\frac{\partial H}{\partial p_\eta}&=&\sum_{s=1}^N\left(\dot{q}_s\frac{\partial p_s}{\partial p_\eta}-\dot{p}_s\frac{\partial q_s}{\partial p_\eta}\right),\label{canonical equations 2 for 1st order for discrete system}
\end{eqnarray}
}
where $\lambda=1,\cdots,\nu$, $\eta=1,\cdots,\mu$, $\nu+\mu=N$, the generalized momenta are defined as
\begin{equation}\label{generalized momenta for discrete system}
p_i=\frac{\partial L}{\partial \dot{q}_i},
\end{equation}
with $L=\int_{-\infty}^\infty ld^D\textbf{r}$ being the Lagrangian,
and the Hamiltonian is obtained by Legendre transformation
as
\begin{equation}\label{Hamiltonian for discrete system}
 H=\sum_{i=1}^n\dot{q}_ip_i-L.
\end{equation}
For the SODS, all the generalized coordinates and the generalized momenta are independent, the new CEH (\ref{canonical equations 1 for 1st order for discrete system }) and (\ref{canonical equations 2 for 1st order for discrete system}) are automatically reduced to the conventional CEH (\ref{qi}) and (\ref{pi}).

We have shown that the FODS can only be expressed by the new CEH,
but do not by the conventional CEH, while the SODS can be done by both the new and the conventional CEHs. We have also shown that the NLSE can be expressed by the new CEH in a consistent way if the propagation coordinate $z$ in the NLSE is considered to be the evolution coordinate.

\section{Application of The new CEH to light-envelope
propagations}\label{application}
Different from the case of the NLSE, the Hamiltonian density of the NNLSE contains the convolution between the response function $R(\mathbf r)$ and the modulus square of the light-envelope $\varphi(\mathbf r,z)$. Following the  procedure in Sec.~\ref{limition on response}, it can be easily proved that the NNLSE can also be expressed with the new CEH in a consistent way if the propagation coordinate $z$ in the model is considered to be the evolution coordinate.
Based on the new CEH, we now introduce a new approach to deal with the nonlinear light-envelope propagations.

We assume the trial solution of the form as
\begin{equation}\label{trial solution}
\varphi(r,z)=q_A(z)\exp\left[-\frac{r^2}{q_w^2(z)}\right]\exp\left[iq_c(z)r^2+iq_{\theta}(z)\right],
\end{equation}
where $q_A,q_{\theta}$ are the amplitude and phase of the complex
amplitude of the light-envelope, respectively, $q_w$ is the width of the light-envelope, $q_c$
is the phase-front curvature, and they all vary with the propagation distance (the evolution coordinate) $z$.
The response function of materials is assumed as
\begin{equation}\label{response function}
R(\textbf{r})=\frac{1}{(\sqrt{\pi}w_m)^D}\exp\left(-\frac{r^2}{w_m^2}\right).
\end{equation}
Inserting the trial solution (\ref{trial solution}) and the response function (\ref{response function}) into the Lagrangian density (\ref{Lagrangian density for NNLSE}),
and performing the integration $L=\int_{-\infty}^\infty ld^D\textbf{r}$ we obtain
\begin{eqnarray}\label{integral Lagrangian}
L&=&-2^{-2-D}\pi^{D/2}q_A^2q_w^{-2+D}(w_m^2+q_w^2)^{-D/2}[-2q_A^2q_w^{2+D}
+2^{D/2}(w_m^2+q_w^2)^{D/2}(4D\nonumber\\
&&+4Dq_c^2q_w^4+Dq_w^4\dot{q}_c+4q_w^2\dot{q}_{\theta})],
\end{eqnarray}
which is a function of generalized coordinates, $q_A,q_w,q_c$ and generalized velocities, $\dot{q}_c,\dot{q}_{\theta}$ (The dot above the variable indicates the derivative with respect to the evolution coordinate $z$), but not an explicit function of the evolution coordinate $z$. Eq.~(\ref{trial solution}) can be understood as a ``coordinate transformation''. Through such a transformation (of course, this is not a  real coordinate transformation in the rigorous sense in mathematics), the coordinate system consist of a set of generalized
coordinates $\varphi$ and $\varphi*$ is transformed to that consist of another set of generalized coordinates $q_A,~q_w,~q_c,$ and $q_{\theta}$, and the Lagrangian density expressed by
 Eq.~(\ref{Lagrangian density for NNLSE}) in the continuous system is transferred to the Lagrangian expressed by Eq.~(\ref{integral Lagrangian}) in the discrete system at the same time via the integration $L=\int_{-\infty}^\infty ld^D\textbf{r}$.

Then the generalized momenta can be obtained by definition (\ref{generalized momenta for discrete system}) as follows
\begin{eqnarray}
p_A&=&p_w=0,\label{p 0}\\
p_c&=&-2^{-2-\frac{D}{2}}D\pi^{D/2} q_A^2 q_w^{2+D},\label{qc momentum}\\
p_{\theta}&=&-\left(\frac{\pi}{2}\right)^{D/2}q_A^2
q_w^D.\label{p_theta}
\end{eqnarray}
The Hamiltonian of the system then can be
determined by Legendre transformation (\ref{Hamiltonian for discrete system})
\begin{eqnarray}\label{Hamiltonian}
H&=&2^{-1-D}\pi^{D/2}q_A^2q_w^{-2+D}(w_m^2+q_w^2)^{-D/2}[-q_A^2q_w^{2+D}\nonumber\\
&&+2^{1+\frac{D}{2}}D(w_m^2+q_w^2)^{D/2}(1+q_c^2q_w^4)],
\end{eqnarray} and can be proved to be a constant, i.e. $\dot{H}=0.$

There are four generalized coordinates and four generalized momenta in the four equations (\ref{p 0}), (\ref{qc momentum}) and (\ref{p_theta}). So the degree of freedom of the set of equations (\ref{p 0}), (\ref{qc momentum}) and (\ref{p_theta}) is four. Without loss of generality, we take $q_c,q_\theta,p_c$ and $p_\theta$ as the independent variables. By solving Eqs.(\ref{qc momentum}) and (\ref{p_theta}), the generalized
coordinates $q_A$ and $q_w$ can be expressed by generalized momenta
$p_c$ and $p_\theta$ as
$
q_A=(-p_\theta)^{1/2}[Dp_\theta/(2\pi
p_c)]^{D/4}$ and
$
q_w=[4p_c/(Dp_\theta)]^{1/2},
$
and inserting this result into the Hamiltonian
(\ref{Hamiltonian}) yields
\begin{eqnarray}\label{Hamiltonian with momenta}
H=-\frac{ D^2 p_{\theta }^2+16 p_c^2 q_c^2}{4 p_c}-\frac{1}{2} \pi
^{-D/2}(\frac{4 p_c}{D p_{\theta }}+w_m^2)^{-D/2}.
\end{eqnarray}
 By use of the canonical equations of Hamilton (\ref{canonical equations 1 for 1st order for discrete system }) and (\ref{canonical equations 2 for 1st order for discrete system}) in the case that $\mu=\nu=2$ and $n=4$ because there are only two independent generalized coordinates and two independent generalized momenta, we can obtain the following four
equations
\begin{eqnarray}
\dot{q}_c&=&\frac{D^2 p_{\theta }^2}{4 p_c^2}-4 q_c^2+\frac{D \pi
^{-D/2} p_{\theta }^2(\frac{4 p_c}{D p_{\theta }}+w_m^2)^{-D/2}}{4
p_c+D p_{\theta } w_m^2},\label{dqc}\\
\dot{q}_\theta&=&-\frac{(4+D) \pi ^{-D/2} p_c p_{\theta } (\frac{4 p_c}{D p_{\theta }}+w_m^2){}^{-D/2}}{4 p_c+D p_{\theta } w_m^2}\nonumber\\
&&-\frac{D^2p_{\theta }}{2p_c}-\frac{D \pi ^{-D/2} p_{\theta }^2
w_m^2 (\frac{4 p_c}{D p_{\theta }}+w_m^2){}^{-D/2}}{4 p_c+D
p_{\theta } w_m^2},\\
\dot{p}_c&=&8 p_cq_c,\label{dpc}\\
\dot{p}_\theta&=&0.\label{dp_theta}
\end{eqnarray}

It can be found that the generalized coordinate $q_{\theta}$ is not contained in the Hamiltonian (\ref{Hamiltonian with momenta}), then $q_{\theta}$ is a cyclic coordinate. It is known that the generalized momentum conjugate to a cyclic coordinate is conserved~\cite{Goldstein-book-05}. Therefore, the generalized momentum $p_{\theta}$  conjugate to the generalized coordinate $q_{\theta}$ is a constant, which
can be confirmed by Eq.(\ref{dp_theta}). In fact, this
represents that the power of the light-envelope,
\begin{equation}\label{power}
P_0=\int_{-\infty}^\infty\left|\varphi\right|^2d^D\textbf{r}=q_A^2(\sqrt{\pi/2}q_w)^D,
\end{equation}
is conservative. Then we can obtain
\begin{equation}\label{amplitude}
q_A^2=P_0(\sqrt{\pi/2}q_w)^{-D}.
\end{equation}
Taking the derivative with respect to $z$ on both sides of
Eq.(\ref{qc momentum}), then comparing it with Eq.(\ref{dpc}), we
can obtain with the aid of Eq.(\ref{amplitude})
\begin{equation}\label{qc}
q_c=\frac{\dot{q}_w}{4q_w}.
\end{equation}
Then by substituting 
 Eq.(\ref{qc}) into the Hamiltonian
(\ref{Hamiltonian}) with the aid of Eq.(\ref{amplitude}), we have
$
H=T+V,
$
where \begin{eqnarray} T&=&\frac{1}{16}DP_0\dot{q}_w^2,\label{kenetic}\\
V&=&\frac{D P_0}{q_w{}^2}-\frac{1}{2} \pi ^{-D/2} P_0^2
\left(w_m^2+q_w^2\right){}^{-D/2}\label{potential energy}
\end{eqnarray}
are the generalized kinetic energy and  the generalized potential of the Hamiltonian system, respectively.

Now we can observe that the dynamics of the light-envelopes in nonlinear media can be treated as problems of small oscillations of a Hamiltonian system about positions of equilibrium from the
Hamiltonian point of view. The equilibrium state of the system described by the Hamiltonian given together by Eqs.~(\ref{kenetic}) and (\ref{potential energy})
corresponds to the soliton solutions of the NNLSE, and can be obtained as the extremum points of the generalized potential of the Hamiltonian system. An equilibrium position is classified as stable if a small disturbance of the system from equilibrium results in small bounded motion about the rest position. The equilibrium is unstable if an infinitesimal disturbance eventually produces unbounded motion~\cite{Goldstein-book-05}. It can be readily seen that when the extremum of the generalized potential is a minimum the equilibrium must be stable, otherwise, the equilibrium must be unstable.
In this sense, therefore, the viewpoint in some
literatures~\cite{Seghete-pra-2007,Picozzi-prl-2011,Lashkin-pla-2007,Petroski-oc-2007}, where
solitons were regarded as the extremum of the Hamiltonian itself rather than the generalized potential of the Hamiltonian system, would be some ambiguous.
Because in those
literatures~\cite{Seghete-pra-2007,Picozzi-prl-2011,Lashkin-pla-2007,Petroski-oc-2007} the trial solution has a changeless
profile (solitonic profile), the state expressed with the solitonic profile is the static state. The kinetic energy of
the static state is zero, and the Hamiltonian is equal to
the potential of the static state. In this connection, the extremum of the
Hamiltonian equals to the extremum of the generalized potential of the static system only in value. Although the soliton solutions obtained in such
literatures~\cite{Seghete-pra-2007,Picozzi-prl-2011,Lashkin-pla-2007,Petroski-oc-2007} are correct, it is more reasonable to consider the soliton solutions of the NNLSE as the extremum points of the generalized potential of the Hamiltonian system.

In order to find the equilibrium position (the soliton solution), letting $\partial V/\partial q_w=0$, we have
\begin{equation}\label{critical power equation}
-\frac{32}{q_w^3}+8 \pi ^{-D/2} P_0
q_w\left(w_m^2+q_w^2\right)^{-1-\frac{D}{2}}=0.
\end{equation}
We can easily obtain the
critical power
 \begin{equation}\label{critical power}
  P_c=\frac{4\pi^{D/2}\left(w_m^2+q_w^2\right)^{1+\frac{D}{2}}}{q_w^4},
\end{equation}
with which the light-envelope will propagate with a
changeless shape. In addition, when $P_0=P_c$, it can be easily obtained that
$
\dot{q}_c=q_c=0,$ which implies that the wavefront
of the soliton solution is a plane.

 Then we elucidate the stability characteristics of the soliton
by analysing the properties of the generalized potential $V$.
Performing the second-order derivative of the generalized
potential $V$ with respect to $q_w$, then inserting the
critical power into it, we obtain
\begin{equation}\label{stability criterion}
\Upsilon\equiv\left.\frac{\partial^2V}{\partial q_w^2}\right|_{P_0=
P_c}=\frac{64 }{q_w^4}\left[2-\frac{2+D}{2
\left(1+\sigma^2\right)}\right],
\end{equation}
where $\sigma=w_m/q_w$ is the degree of nonlocality. The larger is
$\sigma$, the stronger is the degree of nonlocality. When
$\Upsilon>0$, the generalized potential has a minimum, and the soliton is
stable. From Eq.(\ref{stability criterion}) we can obtain the
criterion for the stability of solitons, that is
\begin{equation}\label{stability criterion2}
\sigma^2>\frac{1}{4}(D-2),
\end{equation}
which is, in fact, consistent with the Vakhitov-Kolokolov (VK) criterion~\cite{Vakhitov-qe-75}(for detail, see the footnote~\cite{prove}).

\subsection{The local case}
\begin{figure}[htb]
\centerline{\includegraphics[width=7cm]{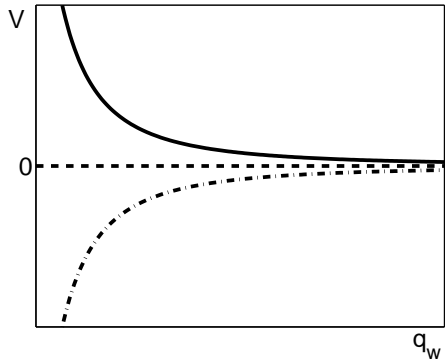}}
\caption{Qualitative plot of the potential $V$ as a function of
$q_w$ for three cases, $P_0<P_c$ (solid curve), $P_0=P_c$
(dashed curve), and  $P_0>P_c$ (dash-dot curve) when
$D=2$.}\label{V}
\end{figure}

When $w_m\rightarrow0$, the response function
$R(\textbf{r})\rightarrow\delta(\textbf{r})$, then the NNLSE will be reduced to the
NLSE (\ref{NLSE}).
In this case, Eqs. (\ref{critical power}) and (\ref{stability criterion}) are reduced to
\begin{equation}
P_c=4\pi^{D/2}q_w^{D-2},
\Upsilon=\frac{32}{q_w^4}(2-D).
\end{equation}
When $D=1$, the critical power is deduced to $P_c=4\sqrt{\pi}/q_w$,
which is consistent with Eq.(42) of Ref.~\cite{Anderson-pra-83}. When $D=2$, the critical power is deduced to $P_c=4\pi$, which is
the same as Eq.(16a) of Ref.~\cite{Desaix-josab-91}. We can
obtain $\Upsilon>0$ when $D<2$, $\Upsilon<0$ when $D>2$, and
$\Upsilon=0$ when $D=2$. So for the local case, the soliton is
stable for (1+1)-dimensional case, but  unstable when $D>2$. It
needs the further analysis for the case of $D=2$ because $\Upsilon=0.$ When
$D=2$, the generalized potential~(\ref{potential energy}) from the
Hamiltonian point of view is deduced to
\begin{equation}\label{local potential energy}
V=\frac{\left(4\pi-P_0\right)P_0}{2\pi q_w^2},
\end{equation}
which
has no extreme when $P_0\neq 4\pi$. When $P_0=P_c=4\pi$, it can be obtained that $V=0$, which is the extreme but not the minimum. So the
(1+2)-dimensional local solitons are unstable. The relation between the potential $V$ and the width $q_w$ of the light-envelope is shown
in Fig.\ref{V}. If the power of the light-envelope equals to the critical power, the potential will be a constant, as can be seen
by dash curve of Fig.(\ref{V}). Without the external disturbance, the light-envelope will stay in its initial state, and keep its
width changeless. But the ideal condition without external disturbances can not exist in fact. If the external disturbance makes the power
larger than the critical power, then the system will evolve towards the lower potential, the beam width will become more and more smaller, and the optical beam will collapse at last, as can be confirmed by the dash-dot curve of Fig.\ref{V}. If the external disturbance makes the power smaller than the critical power, then the system will also evolve towards the lower potential, the beam width will become more and more
larger, and the optical beam will diffract at last, as can be confirmed by the solid curve of Fig.\ref{V}. These conclusions are consist with those of Refs.~\cite{Berge-PR-98,Moll-prl-03,sun-oe-08}.

\subsection{The nonlocal case}
For the nonlocal case, when $D\leq2$, the condition (\ref{stability criterion2}) can be satisfied automatically. That is to say the
(1+1)-dimensional and the (1+2)-dimensional nonlocal solitons are always stable when the response function of the material is a
Gaussian function. It is consistent with the conclusion of Ref.~\cite{Bang-pre-02}. When $D>2$ the solitons can be stable only if the the degree of nonlocality is strong enough that can satisfy the criterion (\ref{stability criterion2}), which is also the same as the result of Ref.\cite{Bang-pre-02}.

\section{two remarks}\label{remarks}
At the end, we make two remarks on the new approach in dealing with the nonlinear light-envelope propagations presented in the paper. Firstly, the new approach is based on the new CEH (\ref{canonical equations 1 for 1st order for discrete system }) and (\ref{canonical equations 2 for 1st order for discrete system}), and we will show that the conventional CEH (\ref{qi}) and (\ref{pi}) will yield contradictory and inconsistent results. Secondly, we will compare our approach with the variational approach, and discuss the differences between them.

\subsection{Contradictory results coming from the conventional CEH }
Here we use the conventional CEH (\ref{qi}) and (\ref{pi}) to deal with the light-envelope propagated in nonlinear media, following the same procedure in Sec.~\ref{application}, and show that the conventional CEH (\ref{qi}) and (\ref{pi}) will give the contradictory and inconsistent results.

Without loss of generality, we only take the NLSE (\ref{NLSE}) with $D=1$ as an example. The NLSE is a special case of the NNLSE when
$w_m$ approaches to zero. Then letting $w_m=0$ and $D=1$ makes Hamiltonian (\ref{Hamiltonian}) reduced into
\begin{equation}\label{Hamiltonian for 1D}
H=\frac{\sqrt{\pi}q_A^2\left[2\sqrt{2}-q_w^2\left(q_A^2-2\sqrt{2}q_w^2q_c^2\right)\right]}{4q_w}.
\end{equation}
Because the Hamiltonian is only the function of the generalized coordinates, the CEH (\ref{qi}), the right hand side of which is the derivative of the Hamiltonian with respect to the generalized momentum, can yield nothing unless $\dot{q}_c=\dot{q}_\theta=\dot{q}_A=\dot{q}_w= 0$. It means the four quantities are all the conserved quantities.  This result coming from the CEH (\ref{qi}) is obviously wrong because such quantities as the amplitude $q_A,$  the width $q_w$ and the phase-front curvature $q_c$ all generally vary with the evolution coordinate $z$ except for the soliton state, and $q_\theta$, the phase of the complex
amplitude of the light-envelope, must be the function of $z$ even for the soliton state.

From the other CEH (\ref{pi}), four equations can be obtained as
\begin{eqnarray}
\dot{p}_c&=&-\frac{\partial H}{\partial q_c}=-\sqrt{2\pi}q_A^2q_cq_w^3,\label{dpc2}\\
\dot{p}_\theta&=&-\frac{\partial H}{\partial q_\theta}=0,\label{dp_theta2}\\
\dot{p}_A&=&-\frac{\partial H}{\partial q_A}=\frac{\sqrt{\pi } q_A\left(-\sqrt{2}+q_A^2q_w^2-\sqrt{2}q_c^2 q_w^4\right)}{q_w},\label{dpA}\\
\dot{p}_w&=&-\frac{\partial H}{\partial q_w}=\frac{\sqrt{\pi}q_A^2 \left(2\sqrt{2}+q_A^2q_w^2-6\sqrt{2}q_c^2q_w^4\right)}{4q_w^2}.\label{dpw}
\end{eqnarray}
 Substitution of the generalized momenta $p_c$ given by Eq.~(\ref{qc momentum}) into Eq.~(\ref{dpc2}) yields the same result as Eq.~(\ref{qc}). Then inserting Eq.~(\ref{qc}) into the Hamiltonian (\ref{Hamiltonian for 1D}) gives out
$
H=\frac{P_0}{16}\left(\dot{q}_w^2+\frac{16}{q_w^2}-\frac{8P_0}{\sqrt{\pi}q_w}\right).
$ The Hamiltonian is the sum of the generalized kinetic energy and the generalized potential
$V(q_w)=\frac{P_0}{2}\left(\frac{2}{q_w^2}-\frac{P_0}{\sqrt{\pi}q_w}\right)$, which is also the same as Eq.~(\ref{potential energy}) when $D=1$ and $w_m=0$. Therefore, the critical power, corresponding to
the extremum point of the generalized potential,
$
P_c=\frac{4\sqrt{\pi}}{q_w}
$
is the same as Eq.~(\ref{critical power}) when $D=1$ and $w_m=0$. It can also be found that Eq.~(\ref{dp_theta2}) is the same as Eq.~(\ref{dp_theta}), which means that the power of the light-envelope
is conservative.
Although the first two equations, Eqs.~(\ref{dpc2}) and (\ref{dp_theta2}), of a set of equations resulting from CEH (\ref{pi}) can give out the correct results, the other two equations, Eqs.~(\ref{dpA}) and (\ref{dpw}), will yield the contradictory and inconsistent results. Let us show as follows.
 Inserting Eq.~(\ref{p 0}) into Eqs.(\ref{dpA}) and (\ref{dpw}) yields
\begin{eqnarray}
P_0&=&\frac{8\sqrt{\pi}}{5q_w},\label{e1}\\
q_c&=&\sqrt{\frac{3}{5}}\frac{1}{q_w^2}.\label{e2}
\end{eqnarray}
Obviously, the two results given by Eqs.~(\ref{e1}) and (\ref{e2}) are both wrong. Under the assumption of the light-envelope with the form of Gaussian-shape given by Eq.~(\ref{trial solution}), the power carried by the light-envelope should be $P_0=\sqrt{\pi/2}q_A^2q_w$ given by Eq.~(\ref{power}), with which Eq.~(\ref{e1}) is contradictory and inconsistent. Eq.~(\ref{e2}) gives the fixed relation between $q_c$ and $q_w$. But the phase-front curvature, $q_c$, should be changed depending upon the state of the light-envelope, especially $q_c$ should be zero for the soliton state, with which Eq.~(\ref{e2}) is inconsistent.

It is no surprise to
obtain such contradictory and inconsistent results from the canonical equations of Hamilton (\ref{qi}) and (\ref{pi}), since both the NNLSE (\ref{NNLSE}) and its complex conjugation can not be derived from the canonical equations of Hamilton (\ref{derivative of q for continuous system}) and (\ref{derivative of pi for continuous system}) as stated in Sec.~\ref{ceh for 1st system}.

\subsection{Our approach vs the variational approach: same and different}\label{difference}
As mentioned above, our approach presented in this paper is based on the canonical
equations of Hamilton (the Hamiltonian formulation), while the variational approach~\cite{Anderson-pra-83} is based on the Euler-Lagrange equations (the Lagrangian formulation). Although the same point of the two approaches is to first compute the Lagrangian of the system by using a suitably chosen trial function,
they are in essence two parallel methods
because the Hamiltonian formulation and the Lagrangian formulation are two parallel theory frameworks in the classical
mechanics.

The most important concept in our approach is the ``potential''. The potential given by Eq.~(\ref{potential energy}) is the real ``potential'' of the system that a single
light-envelope propagates in nonlocal nonlinear media modeled by the NNLSE. It is not, of course, the potential of the narrow-sense mechanical system, but does be the potential in the frame of the Hamiltonian theory, that is, the potential of the Hamiltonian system. In other word, it is the potential from the Hamiltonian point of view. Looking back to the variational approach, we can observe that although the ``potential'' was also introduced [see, Eqs.~(28) and (29) in Ref.~\cite{Anderson-pra-83}], it is just a mathematically
equivalent potential in the sence that the evolution of the width of the light-envelope can be analogous to that of a particle in a potential well, rather than the real ``potential'' of the system.


\section{Conclusion}\label{conclusion}
We introduce a new approach, based on the new canonical
equations of Hamilton found by us recently, to analytically obtain the approximate solution of the nonlocal nonlinear Schr\"{o}dinger equation and to analytically discuss the stability of the soliton.
For the single light-envelope propagated in nonlocal nonlinear media modeled by the NNLSE, the Hamiltonian of the system
can be constructed as the sum of the generalized kinetic energy and the generalized potential. The extreme point of the
generalized potential corresponds to the soliton solution of the NNLSE. The soliton is stable when the generalized
potential has the minimum, and unstable otherwise. In addition, we give the rigorous proof of the equivalency between the NNLSE and the Euler-Lagrange equation on the premise of the response function with even symmetry.

%
\section*{ACKNOWLEDGMENTS}
This research was supported by the National Natural Science
 Foundation of China, Grant Nos.~11274125 and 11474109.


\begin{thebibliography}{99}
\bibitem{prove} To avoid confusion,
we make the notations used in Ref.\cite{Bang-pre-02} consistent with
those in the paper, where we use $w_m$ and $q_w$ to represent the
width of the response function and the light-envelope instead of
$\sigma,\beta$, respectively. In Ref.\cite{Bang-pre-02},
$P_s=(\pi/2)^{D/2}\alpha^2q_w^D$, with
$\alpha^2=(\lambda+D/q_w^2)(2+2w_m^2/q_w^2)^{D/2}$ and
$q_w^2=\left[4-D+\sqrt{(4-D)^2+16\lambda w_m^2}\right]/(2\lambda)$,
where $P_s$ is the soliton power, and $\lambda$ is the propagation
constant of solitons. According to the VK criterion, when
$dP_s/d\lambda>0$ the soliton become linearly stable.
We can derive
$\sigma^2>\frac{1}{4}(D-2)$ from the VK criterion $dP_s/d\lambda>0$,
where $\sigma=w_m/q_w$ represents the degree of nonlocality.
\end{thebibliography}
\end{document}